\documentclass[aps,twocolumn,prl,superscriptaddress]{revtex4}
\usepackage[dvips]{graphics}
\usepackage{graphicx}
\usepackage{amsfonts}
\usepackage{amssymb}
\usepackage{amsmath}
\usepackage{subfigure}

\begin{document}

\title{Aging dynamics and the topology of inhomogenous networks}
\author{R. Burioni}
\affiliation{Dipartimento di Fisica and INFN, Universit\`a di Parma, 
Parco Area delle Scienze 7/A, I-423100 Parma, Italy}
\author{D. Cassi}
\affiliation{Dipartimento di Fisica and INFN, Universit\`a di Parma, 
Parco Area delle Scienze 7/A, I-423100 Parma, Italy}
\author{F.Corberi}
\affiliation{Dipartimento di Fisica ``E.R.Caianiello'' and CNR-INFM, Universit\`a 
di Salerno, 84081 Baronissi (Salerno), Italy}
\author{A. Vezzani}
\affiliation{Dipartimento di Fisica and CNR-INFM, Universit\`a di Parma, 
Parco Area delle Scienze 7/A, I-423100 Parma, Italy}

 \date{\today}

\begin{abstract}
We study phase ordering on networks and we establish a relation between the exponent $a_\chi$ of 
the aging part of the integrated autoresponse function $\chi _{ag}$ and the topology 
of the underlying structures. 
We show that $a_\chi >0$ in full generality on networks which are above the lower 
critical dimension $d_L$, i.e. where the corresponding statistical model  
has a phase transition at finite temperature. 
For discrete symmetry models on finite ramified structures with $T_c = 0$, which are at 
the lower critical dimension $d_L$, we show that  $a_\chi$ is expected to vanish.
We provide numerical results for the physically interesting case
of the $2-d$ percolation cluster at or above the percolation threshold, i.e. at or above $d_L$,
and for other networks, showing that the value of $a_\chi $ changes according to our hypothesis.
For $O({\cal N})$ models we find that the same picture holds in the large-${\cal N}$ limit
and that $a_\chi$ only depends on the spectral dimension of the network.

\end{abstract}


\maketitle

\draft
\def\be{\begin{equation}}
\def\ee{\end{equation}}
\def\bfi{\begin{figure}}
\def\efi{\end{figure}}
\def\bea{\begin{eqnarray}}
\def\eea{\end{eqnarray}}

In recent years, networks and graphs have been successfully applied  
to very different disciplines~\cite{barabasi}. 
This extreme flexibility is not surprising, 
since they are the most general way of representing a set of elements 
connected pairwise by some kind of relation. 
In social sciences, 
computing, psychology and economy, graphs represent communication systems, 
social relationships, biological interactions and statistical models 
of algorithms.   
In physics, chemistry 
and biology, they are extensively used as models for real complex 
and inhomogeneous structures, such as  disordered 
materials, percolation clusters, glasses, polymers, biomolecules.
Networks representing real physical structures are
constrainted to be embeddable in a finite dimensional space
and to have bounded coordination number, forming an interesting class
of structures called "physical graphs" \cite{rassegna}.   
 
Basic physical equilibrium properties of models defined on physical graphs, 
among which critical behaviors, depend crucially on the topological features 
of the network. In particular, the relation between 
large scale topology and critical properties
is well understood 
on a simple class of network, i.e. regular lattices, where 
the existence of phase transitions,
the value of critical exponents and many other 
equilibrium quantities 
depend on a topological parameter, the Euclidean dimension $d$ of the lattice.
As far as systems with continuous symmetry are concerned,
such as $O({\cal N})$ models,
this general picture can be extended to networks. Their topology can be 
completely described by the adjacency matrix $A_{i,j}$, with $A_{i,j}=1$ if $i$ and $j$ are connected by a link and $A_{i,j}=0$ otherwise, but the relevant large scale topological information is encoded in the Laplacian matrix $\Lambda_{i,j}= \delta_{i,j} - z_i$, where $z_i = \sum_j A_{i,j}$ is the number of neighbours of site $i$.
Indeed, the density $\rho(\lambda)$ of eigenvalues  
of $\Lambda$, $\rho(\lambda)\sim \lambda^{(d_s/2)-1}$ for $\lambda \to 0$ defines
the "spectral dimension" $d_s$ of the graph.
The spectral dimension univoquely determines the existence of phase transitions
\cite{fss} and controls critical behaviour \cite{sferico}, much in the 
same way as the Euclidean dimension $d$ does on usual translation 
invariant lattices.
In systems with a discrete symmetry,
although a unique topological indicator, analogous to $d_s$, is not known, the fundamental 
role of topological and connectivity
properties in determining equilibrium and critical behavior 
has been pointed out~\cite{aharony}.

The picture is much less clear for out of equilibrium processes. Although it is known 
that dynamical processes are influenced by the topology of the network, a direct description 
of this relation is largely incomplete. This is the case for
the phase ordering dynamics, occurring in systems gradually evolving from an
homogeneous phase to a state with two or more coexisting ordered phases, through
formation and growth of segregated domains~\cite{bray}. 
Initally investigated in physics, phase ordering also occurs in other fields where
the topology is of primary importance, as in biology, where it describes spreading
of tissues \cite{pnas}, and in economic and social networks,
where it has been used as a model for propagation of opinions and technologies
\cite{cmv}.
A deeper understanding of the relation between the dynamical
process and the topology of the network in these frameworks would be of great interest.

In the simpler case of regular lattices, the influence of the 
Euclidean dimension on dynamical exponents has been ascertained
in phase ordering in spin systems~\cite{bouchaud}.
The generalization of such models to networks is provided by the 
Hamiltonian $H\{ \vec \sigma \} = \sum_{i, j}A_{i,j} \vec{\sigma_i}\cdot \vec{\sigma_j}$,
where the indices $i$ and $j$ run over the $N$ sites of the graph,
and $\vec{\sigma_i}$ represents an $\cal N$-dimensional unit vector defined at the site $i$.
For ${\cal N}=1$, $H$ describes
the discrete Ising model and for ${\cal N}>1$
the $O({\cal N})$ models, with continuous symmetry. 
Phase ordering dynamics occurs when
the system, initially at thermal equilibrium at high temperature, 
is quenched to a temperature where the symmetry is broken and more
phases coexist. In this scenario, an interesting physical quantity
is the integrated autoresponse function $\chi (t,s)$ describing the effect of a 
small perturbation acting on the system from time $s$ to $t$: 
$\chi (t,s)=\int _s ^t R(t,t')dt'$,
where $R(t,t')=N^{-1} \sum _i \delta \langle \vec \sigma _i (t)\cdot 
\vec n_i \rangle /\delta |\vec h_i(t')| \vert _{\vec h_i(t')=0}$
is the average linear response function associated to an impulsive magnetic field
switched on in $i$ at $t'$~\cite{noi} ($\vec n_i = \vec h_i/ |\vec h_i|$
and  $\langle \dots \rangle$ means ensemble averages). 
Interestingly, $\chi (t,s)$, can be splitted into a stationary and an 
aging term~\cite{bouchaud}
$\chi (t,s)=\chi _{st}(t-s)+\chi _{ag}(t,s)$, and the aging part features a
scaling behaviour with a characteristic exponent $a_\chi$
\be
\chi _{ag}(t,s)=s^{-a_\chi}f(t/s).
\label{scalchi}
\ee
{\sf In phase ordering on lattices~\cite{noi}}
\be
f(x)\sim x^{-a_\chi }\quad \mbox{for} \quad x\gg 1, 
\label{largex}
\ee
and $a_\chi $ is related to $d$ by
\be
a_\chi =\theta \frac{d - d_L}{2}
\label{expa}
\ee
for $d<d_U$, while $a_\chi =\theta$ for $d \ge d_U$. 
Here $d_L$ is the lower critical dimension, $d_U=3$ or $d_U=4$
for discrete and continuos symmetry models respectively,
and $\theta$ is the exponent regulating
the time decay of the topological defects density $\rho \propto t^{-\theta }$, which does not depend on $d$. 
By means of Eq.~(\ref{expa}), a non-equilibrium exponent, $a_\chi$,  
is related to the topology of the underlying lattice, through $d$.  
This relation implies that
$a_\chi>0$ when the system is above $d_L$, i.e. when a phase transition 
at finite temperature $T_c$ occurs, 
$a_\chi = 0$ at $d_L$ and $a_\chi < 0$ below $d_L$.

These results, holding in the simple case of regular lattices, 
suggest that 
non-equilibrium dynamics of statistical models 
could be related to important topological properties of networks.
With the aim of testing and investigating this hypothesis,
we study the response function exponent $a_\chi $ 
on generic physical networks, showing that i)
$a_\chi >0$ is expected in full generality on networks
supporting phase transitions at finite $T_c$, 
i.e. above $d_L$; 
ii) for continuous symmetry $O({\cal N})$ models,
in the soluble large-${\cal N}$ limit 
on graphs
the same expression~(\ref{expa}) is found, with $d_s$ occurring 
in place of $d$.
This fits with the aforementioned general picture that, 
for continuous symmetry models 
$d_s$ is the topological indicator replacing $d$ on graphs;
iii) for discrete symmetry models,
we provide an argument showing that
$a_\chi =0$ is expected on finitely ramified networks (FRNs)\cite{aharony}, 
which represents the most general class of structures without 
phase transitions at finite $T_c$.
Although in this case a topological
parameter playing the role of $d_s$ is not known and, therefore,
a generalization of Eq.~(\ref{expa}) is not straightforward as
for continuous symmetries, this result conforms with Eq.~(\ref{expa})
and provides, via the existence of phase transitions,
a relation between the positivity of $a_\chi$ and the topological 
feature of the network. 
Our results are complemented by extensive
numerical simulations of different discrete symmetry models on a representative set of
graphs finding $a_\chi =0$ or $a_\chi >0$
depending on the absence/presence of phase transitions.
In particular, we provide results for the physically interesting case
of the $2-d$ percolation cluster at or above the percolation threshold,
i.e. at or above $d_L$, 
showing that the value of $a_\chi$ changes according to our hypothesis.
Taking advantage
from this, $a_\chi $ could be conveniently used to investigate
the phase diagram of statistical models on general discrete structures. 

Let us start showing that i) $a_\chi >0$ is expected for models with
a phase transition at finite $T_c$. 
Reparametrizing the two time dependence of $\chi (t,s)$ in terms 
of the autocorrelation function 
$C(t,s)=1/N \sum_{i=1}^N \langle \vec \sigma _i(t) \cdot \vec \sigma _i(s)\rangle$ ~\cite{cugliakurc}, 
one obtains the parametric form
$\widehat \chi (C,s)$, which is related to the structure of the equilibrium state through \cite{Franz98}
\be
\left . -T\lim _{s\to \infty }\frac{d^2\widehat \chi (C,s)}
{dC^2}\right \vert _{C=q}=P_{eq}(q) 
\label{theorem}
\ee
where $P_{eq}(q)=Z^{-2}\sum _{\vec \sigma, \vec \sigma '}
\exp [-\beta (H\{ \vec \sigma \}+H\{ \vec \sigma '\})]
\delta [Q(\vec \sigma,\vec \sigma ')-q]$  is the equilibrium
probability distribution of the overlaps between two configurations $\{\vec \sigma\}$ and $\{\vec \sigma '\}$,
$Q(\vec \sigma,\vec \sigma ')=(1/N)\sum _i \vec \sigma _i \cdot \vec \sigma _i'$, $Z$ being the
partition function. 
General considerations can be drawn on the relation between 
$\chi _{ag}$ and the critical behavior of the corresponding statistical model.
Let us consider a 
system with $T_c>0$ quenched to
$0<T<T_c$. In this case the form of 
$P_{eq}(q)$ implies~\cite{lungovecchio} that $\chi _{ag}$
must vanish asymptotically. Then, from Eq.~(\ref{scalchi}),
one has $a_\chi>0$.
Even if for quenches to $T=0$ 
Eq.~(\ref{theorem}) does not give informations on  
$\chi _{ag}$, since dynamical exponents are known~\cite{bray} not to depend on $T$ ~\cite{bray}, 
$a_\chi>0$ must be found in the whole interval $0\leq T< T_c$. 
These considerations being completely general, they apply to networks as well,
and $a_\chi >0$ for quenches to $T<T_c$ in systems with finite $T_c$.

Let us now consider point ii).
In the large-$\cal N$ limit,  the order parameter $\phi_i$ can be expanded in the base of the eigenvectors of the Laplacian operator and the dynamical evolution of the projection
$\hat \phi_k$ of the order parameter
on the $k$-th eigenvector of the Laplacian operator obeys~\cite{umberto}
\be
{\partial \hat \phi _k(t)  \over \partial t} = -\left [ \lambda_k + I(t)\right ] 
\hat \phi _k + \hat \eta_k(t)  
\label{eqev_k}
\ee
where $\lambda_k$ is the $k$-th eigenvalue of $\Lambda$, 
$I(t)= 1/N \sum_i^N \langle \phi _i^2\rangle - 1$ and $\hat \eta _k(t)$ is the 
thermal noise. 
Eq.~(\ref{eqev_k}) is formally linear and
it can be solved in the large time domain.
Proceeding as in \cite{fedeon} to compute $\chi _{ag}(t,s)$
for a quench below $T_c$ and extracting the leading term, 
corresponding to the low eigenvalue behaviour of $\rho(\lambda)$,
we obtain the forms~(\ref{scalchi},\ref{largex}), with 
$a_{\chi}=(d_s - 2)/2$ for $d_s<4$ while $a_{\chi}=1$ for $d_s\ge 4$.
This expression of $a_\chi$ is analogous to Eq.~(\ref{expa}),
with the Euclidean dimension 
replaced by $d_s$, $d_L=2$, $d_U=4$ 
and $\theta =1$, as appropriate to vectorial models~\cite{bray}.
This also implies that when $d_s >d_L$, i.e. $T_c >0$ \cite{fss},
$a_\chi>0$, consistently with our previous argument. 
Interestingly, 
most networks representing real physical structures 
have $d_s < 2$ \cite{esempids} and, therefore, $a_\chi<0$. This would
imply a diverging response.

\begin{figure}
    \centering
   \rotatebox{0}{\resizebox{.45\textwidth}{!}
     {\includegraphics{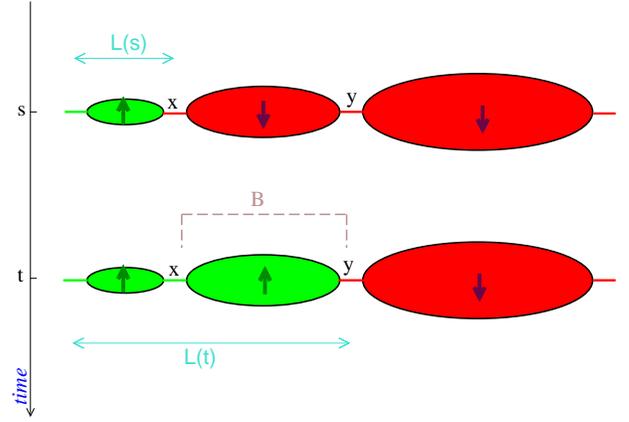}}}
    \caption{Schematic representation of a FRN.}
\label{parma}
\end{figure}

Let us now turn to discrete symmetry systems without phase 
transition, i.e. at $d_L$. 
In this case $T_c=0$ and aging dynamics can only be observed at $T=0$ 
where Eq.~(\ref{theorem}) does not provide 
a positivity constraint on $a_\chi$. 
For instance, the exact solution of the kinetic Ising 
chain~\cite{lip} gives $a_\chi=0$ for $T\to 0$. 
In the following we introduce a topological argument showing that the same
result $a_\chi=0$ holds for coarsening systems 
on FRNs. On these structures, each arbitrary
large part can be disconnected cutting a finite number of links,
as those marked $x$ or $y$ in the 
representation of Fig.~\ref{parma}.
For FRNs  $T_c=0$ \cite{aharony}, 
and even if the general condition for  $T_c=0$ 
for discrete symmetry models is not known, the 
available results indicate that large scale connectivity is the 
only relevant feature and that infinite ramification on scale invariant
structures implies $T_c>0$ ~\cite{aharony}.  

Let us consider phase ordering on a FRN.
We assume a no bulk flip rule (NBF), where spins in the
bulk of ordered domains cannot flip. It has been shown~ \cite{teff} that
this dynamics isolates the aging part of $\chi $ leaving other properties
of the system unchanged. 
Switching on a random magnetic field with expectations 
$\overline {h_i}=0$, $\overline {h_i h_j}=h^2 \delta _{ij}$ from time $s$ 
onwards, $\chi _{ag}$ can be computed as~\cite{noi,lungovecchio}
\be
\chi (t,s)=\lim _{h\to 0}\frac{1}{Nh^2}\sum _i ^N \overline {\langle {\sigma }_i(t)  {h}_i \rangle},
\label{chicampo}
\ee
where $\sigma_i=\pm 1$ and
$\overline {\cdots }$ denotes an average over the realizations of $h_i$. 
Let us focus on a domain of, say, 
up spins whose interface lies at time $s$ on the 
link denoted $x$, and is subsequently found in $y$ at time $t$. 
The domain's size grows from $L(s)$ to $L(t)$.
The number of spins 
in the region $B$ between $x$ and $y$
is $n_B\propto [L(t)-L(s)]^{d_c}$,
where $d_c$ is the connectivity dimension of the network~\cite{rassegna}.  
We want to estimate the increase of the response $\chi _{ag}^{sing}(t,s)$ 
associated to the displacement of this single interface from $x$ to $y$.
Eq.~(\ref{chicampo}) gives
$\chi _{ag}^{sing}(t,s)=-(Nh^2)^{-1}\overline {\sum _y E_h P_h(y,t)}$,
where $P(y,t)$ is the probability to find the interface in $y$
at time $t$ and $E_h=-\sum _{i\in B}h_i\sigma _i=-\sum _{i\in B}h_i$ 
is the variation of
magnetic energy due to its displacement from $x$ to $y$.
Now we make the assumption that the correction to the unperturbed probability
$P_0(y,t)$ is in the form of a Boltzmann factor 
$P_h(y,t)= P_0(y,t)\exp (-\beta E_h)\simeq P_0(y,t)[1-\beta E_h]$.
Then $\chi _{ag}^{sing}(t,s)=
-(Nh^2)^{-1}\overline {\sum _y E_h [1-\beta E_h]P_0(y,t)}$. 
Since the linear term in $E_h$ vanishes by symmetry
we find $\chi _{ag}^{sing}(t,s)=
\beta (Nh^2)^{-1}\overline {\sum _y E_h^2 P_0(y,t)}=\beta (Nh^2)^{-1}\overline {\langle E_h^2\rangle}$.
Recalling the expression of $E_h$ and the expectations of $h_i$
one has $\overline {\langle E_h^2\rangle}\simeq h^2 n_B$.
Then  $\beta ^{-1}\chi _{ag}^{sing}(t,s)\propto N^{-1} [L(t)-L(s)]^{d_c}$.
The total aging response of the system at time $t$ is obtained multiplying $\chi _{ag}^{sing}(t,s)$
by the number $n_i(t)=N L(t)^{-d_c}$ of interfaces present in the system.
Then, for $t\gg s$ one has
$\beta ^{-1}\chi _{ag}(t,s)=n_i(t)\beta ^{-1}\chi _{ag}^{sing}(t,s)\sim const.$
Recalling Eq.~(\ref{largex}), this implies $a_\chi=0$. Putting togheter i) and iii) we conclude that 
$a_\chi >0$ or $a_\chi=0$ is expected in the phase ordering
of discrete symmetry systems on networks with $T_c>0$ or $T_c=0$ respectively.

In order to check this general result, for the Ising model 
we have determined numerically 
$a_\chi$ in a representative class of structures. 
We have used the NBF to extract the aging part of two time functions.
Notice that the NBF suppresses bulk fluctuations and hence rises 
the critical temperature~\cite{oliveira},
allowing one to avoid numerically demanding low temperature regions.
The parametric plot of $\widehat \chi _{ag} (C_{ag},s)$ versus $C_{ag}$,
is well suited to discriminate between $a_\chi =0$ and $a_\chi >0$. 
In fact, since $C_{ag}(t,s)=g(t/s)$~\cite{bray}, with $g$ a
monotonously decreasing function, from Eq.~(\ref{scalchi})
one has j) collapse of $\widehat \chi _{ag}(C_{ag},s)$ on a single master-curve, 
for $a_\chi=0$ or jj) lowering of the curves increasing $s$, for $a_\chi >0$.

\begin{figure}
    \centering
    \rotatebox{0}{\resizebox{.45\textwidth}{!}{\includegraphics{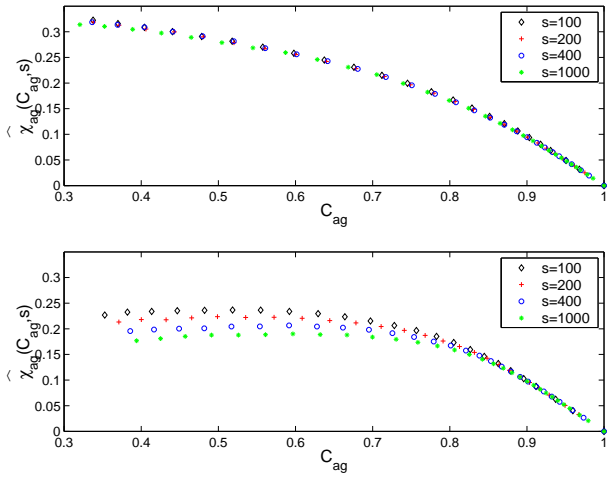}}}
    \caption{$\widehat \chi _{ag}(C_{ag},s)$ for the Ising model quenched to $T=1.25$ 
      on the site-percolation cluster 
      at $p=p_c=0.407$ (upper panel) and at $p=0.2$ (lower panel). 
      The cluster is built on a $1200^2$ 
      square lattice.
      Data are averaged over
      $100$ realizations.}
    \label{figpercolation}
\end{figure}

\begin{figure}
    \centering
    
   \rotatebox{0}{\resizebox{.45\textwidth}{!}{\includegraphics{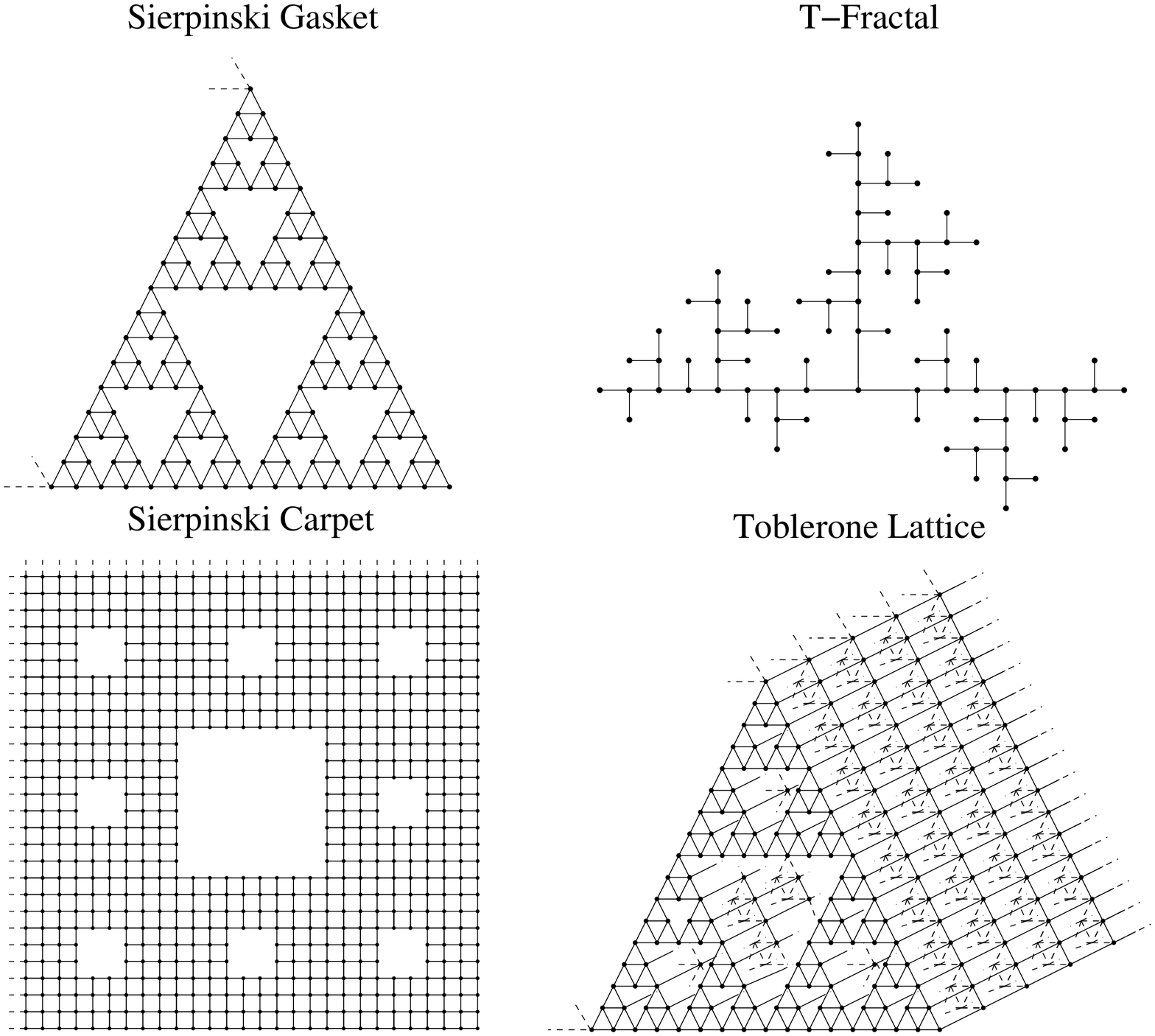}}}
    \caption{The discrete networks considered in simulations: 
      265722 sites Sierpinski gasket, 531442 sites T-fractal, 1048576 sites Sierpinski 
      carpet and 5078988 sites Toblerone lattice.}
\label{fig_net}
\end{figure}

\begin{figure}
    \centering
    
   \rotatebox{0}{\resizebox{.45\textwidth}{!}{\includegraphics{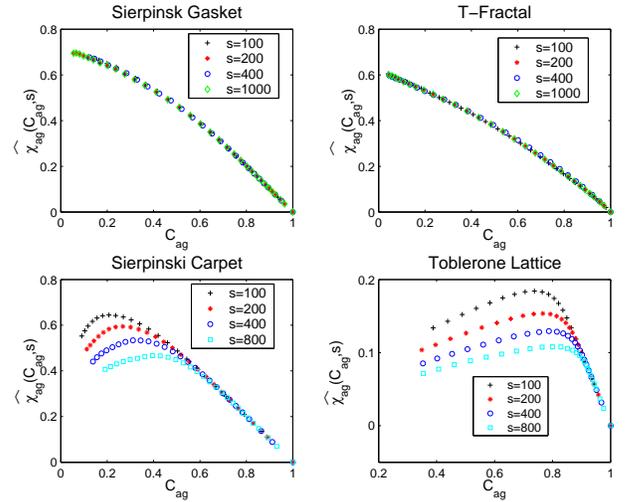}}}
    \caption{$\widehat \chi _{ag}(C_{ag},s)$ for the Ising model at $T=3.0$. 
      On FRNs 
      $a_\chi \simeq 0$; for the Sierpinski Carpet and the Toblerone lattice
      $a_\chi \simeq 0.17$ and $a_\chi \simeq 0.25$.}
\label{fig_frat}
\end{figure}

Coarsening on a percolation cluster is suited to describe
our results because the bond probability $p$ distinguishes a situation with
$T_c=0$ at the percolation threshold $p=p_c$~\cite{stauffer}, from those with $T_c>0$ when $p>p_c$. 
Fig.~\ref{figpercolation} shows that indeed for the Ising model one
has $a_\chi=0$ at $p=p_c$ and $a_\chi>0$ for $p>p_c$.
Furthermore, we have performed analogous simulations on two FRNs with
$T_c=0$, the Sierpinski gasket and the T-fractal (Fig.~\ref{fig_net}),
where $a_\chi=0$ with great accuracy (Fig.~\ref{fig_frat}). 
The situation is different for 
infinite ramification,  with $T_c>0$, 
e.g. on the Sierpinski carpet and the Toblerone lattice \cite{toble}
$a_\chi>0$ (Fig.~\ref{fig_frat}).
Similar results are found 
for different discrete symmetry models with non-conserved order parameter~\cite{bray}, such as 
the 3-state Potts model, confirming the generality of our result.
Interestingly, in all our simulations, the  value of $a_\chi$ 
has been found to be extremely stable and independent of the temperature at 
which the simulation was performed, even when other asymptotic quantities
were strongly fluctuating. This suggests that $a_\chi$ should be related to a
true universal characterization of the process.    

In conclusion, we have shown that, in all the cases considered,
statistical models on networks above $d_L$
are characterized by a positive 
non-equilibrium exponent $a_\chi $, 
while structures at $d_L$
have $a_\chi = 0$.     
For continuous symmetry systems, our large-$N$ calculation shows
that, besides this, the whole dependence of $a_\chi $ on the network topology 
can be expressed as for regular lattices, namely Eq.~(\ref{expa}), 
with the spectral dimension $d_s$ replacing
the Euclidean dimension $d$. 
Our results provide an evidence for a relationship 
between non-equilibrium 
kinetics and large scale topology on general networks
and suggest that the same topological 
features of the networks 
determine critical behavior and non-equilibrium exponents of phase ordering.

\end{document}